# Development of As-Se tapered suspended-core fibers for ultra-broadband mid-IR wavelength conversion

E.A. Anashkina[1,2,*], V.S. Shiryaev[3], M.Y. Koptev[1], B.S. Stepanov[3], S.V. Muravyev[1]

[1] Institute of Applied Physics of the Russian Academy of Sciences, Nizhny Novgorod, Russia
[2] University of Nizhny Novgorod, Nizhny Novgorod, Russia
[3] G.G. Devyatykh Institute of Chemistry of High-Purity Substances of the Russian Academy of Sciences, Nizhny Novgorod, Russia
*elena.anashkina@gmail.com

**Highlights**:
- As-Se tapered suspended-core fibers were designed for mid-IR SC generation.
- Parameters of the produced fibers were studied experimentally and numerically.
- SC in the 1-10 µm range with 150-fs 100-pJ pump at 2 µm was obtained numerically.
- Experiments on spectral broadening of pump pulses at 1.57 µm were performed.

**Abstract**
We designed and developed tapered suspended-core fibers of high-purity $As_{39}Se_{61}$ glass for supercontinuum generation in the mid-IR with a standard fiber laser pump source at 2 µm. It was shown that microstructuring allows shifting a zero dispersion wavelength to the range shorter than 2 µm in the fiber waist with a core diameter of about 1 µm. In this case, supercontinuum generation in the 1-10 µm range was obtained numerically with 150-fs 100-pJ pump pulses at 2 µm. We also performed experiments on wavelength conversion of ultrashort optical pulses at 1.57 µm from Er: fiber laser system in the manufactured As-Se tapered fibers. The measured broadening spectra were in a good agreement with the ones simulated numerically.

**1 Introduction**
The mid-infrared (IR) supercontinuum (SC) laser sources have important applications in both basic science and industry. These light sources are often based on ultra-broadband nonlinear spectral conversion in fluoride, tellurite, and chalcogenide glass (ChG) fibers [1-4]. Among different glass compositions, the ChGs have the highest nonlinearity and the broadest transparency ranges which allow operating at wavelengths where fluoride and tellurite glasses cannot be transparent [1]. To date, the broadest SC demonstrated in ChG fibers extends from ~2 to 16 µm [4]. It is well-known that not only the nonlinearity and transparency range impact on spectral broadening; the zero-dispersion wavelength (ZDW) location with respect to a central pump wavelength is also of great importance for SC width [5,6]. Operation in the anomalous dispersion regime often allows obtaining a broader spectrum in comparison with normal dispersion. In this pumping scheme, SC generation is dominated by the soliton dynamics. To generate ultra-broadband SC in solid step-index ChG fibers having ZDW near the material ZDW (which is significantly beyond 4 µm, for example, 4.81 µm for $As_2S_3$, 7.5 µm for $As_2Se_3$, and >7.5 µm for Te-based ChG [1]), mid-IR optical parametric oscillators and amplifiers have been employed [7, 8]. However, if it is desirable to use a pump source at shorter wavelengths than material ZDW, microstructured or/and tapered fibers with blue-shifted ZDW can be implemented. Suspended-core fiber (SCF) geometry is widespread for dispersion management. So, SC spanning 1.7-7.5 µm has been obtained in As-Se SCF with ZDW of 3.5 µm pumped at 4.4 µm [9].

In this paper, the design, development, and manufacturing of tapered SCF with a thin waist core diameter of about 1 µm of the high-purity As-Se glass are reported. The As-Se glass has a broad spectral transparency range (0.85-17.5 µm for absorption coefficient level of 1 $cm^{-1}$), a potentially low optical loss in the 4-7 µm range, a large third-order nonlinear index ($n_2 \sim 2 \cdot 10^{-17}$ $m^2/W$), low phonon energy (250 $cm^{-1}$), low thermo-optical coefficients ($dn/dT \sim (1-5) \cdot 10^{-6}$ $°C^{-1}$), high chemical stability, and excellent resistance to atmospheric moisture. The tapered SCF from this glass is developed for wavelength conversion to the mid-IR of ultrashort signals from a standard *near-IR* fiber laser system in contrast to most experimental and theoretical works dealing with mid-IR pump sources. Tapering is an important optical fiber reprocessing process that enables engineering the total dispersion from normal at the input thick end to anomalous in the fiber waist at a pump wavelength. Using a relatively thick core at the input fiber end leads to increasing efficiency of pump pulse launching. Besides, tapering significantly increases the nonlinear parameters, which beneficially impacts on nonlinear wavelength conversion. We present experimental and theoretical study of produced samples parameters as well as numerical simulation of SC generation in SCF pumped at 1.57 and 2 µm. Experiments on signal spectral broadening in tapered SCF pumped at 1.57 µm are performed. Challenges and prospects of using designed As-Se tapered SCF for mid-IR SC sources are also discussed.

**2 Results**
*2.1 SCF design*
To prepare the preform for 3-hole SCF, the $As_{39}Se_{61}$ glass composition was chosen. The $As_{39}Se_{61}$ glass has a higher stability against crystallization as compared to $As_{40}Se_{60}$ composition [10], allowing to carry out a two-step fiber drawing process without a significant increase in addition optical losses.

The $As_{39}Se_{61}$ glass was prepared by direct melting of as-received specially pure starting elements in an evacuated silica-glass ampoule in a rocking furnace using the chemical-distillation method of melt purification [11, 12]. The starting Se with 6N purity additionally purified by double vacuum distillation, and the arsenic with 5N purity additionally purified by vacuum sublimation were used. The batches of purified selenium and arsenic were loaded into a silica-glass reactor by vacuum evaporation from intermediate ampoules [13]. To purify the As-Se glass from oxygen and hydrogen impurities, the Al (700 ppm) and the $TeCl_4$ (2000 ppm) were loaded into the reactor in an inert atmosphere box. Then, the charge in the evacuated silica-glass ampoule was melted in a rocking furnace at a temperature of 800°C for 8 h. During melting, the impurity oxygen reacting with Al formed a non-volatile $Al_2O_3$, and chlorine from $TeCl_4$ reacted with impurity hydrogen to form volatile HCl. Then, the As-Se melt was purified from these limited impurities by triple vacuum distillation, with subsequent homogenization of distillate at 750°C for 5 h, and air-quenching, annealing at 175°C for 1 h, and cooling to room temperature were performed. The rod-shaped glass sample was prepared with a diameter of 16 mm and a length of 150 mm. Then the glass sample was cut in the form of a rod with a length 80 mm and in the form of the disc of 2 mm thickness with parallel-plate polished surface plates.

The glass composition after the final distillation and homogenization examined by energy-dispersive X-ray microanalysis was $As_{39.0\pm0.2}Se_{61.0\pm0.2}$. The differential scanning calorimetry (DSC) analysis of $As_{39}Se_{61}$ glass showed that the glass transition temperature ($T_g$) determined at a heating rate of 10 K/min was 177±1°C. The DSC curve is characterized by the absence of crystallization peaks. The XRD study of $As_{39}Se_{61}$ glass investigated using X-ray diffractometer XDR-6000 Shimadzu (CuKα-radiation) did not reveal the presence of crystalline phases within the sensitivity of the method (<0.3 vol.%). The as-prepared $As_{39}Se_{61}$ glass was analyzed by means of IR-spectroscopy using the Fourier transform IR spectrometer IRP Prestige–21 (Shimadzu, Japan) in the 7000-350 $cm^{-1}$ spectral range.

The transmission spectra of an $As_{39}Se_{61}$ glass sample with thickness 2 mm and 80 mm are presented in Fig. 1. The transmission spectra of this glass do not show intensive impurity absorption bands, with the exception of the weak absorption band of Se-H groups (2180 $cm^{-1}$) in the sample with 80 mm thickness. In the long-wavelength region of the transmission spectrum, there are fundamental multiphonon stretching absorption bands at 480 and 720 $cm^{-1}$ due to vibration of As-Se bonds [14].

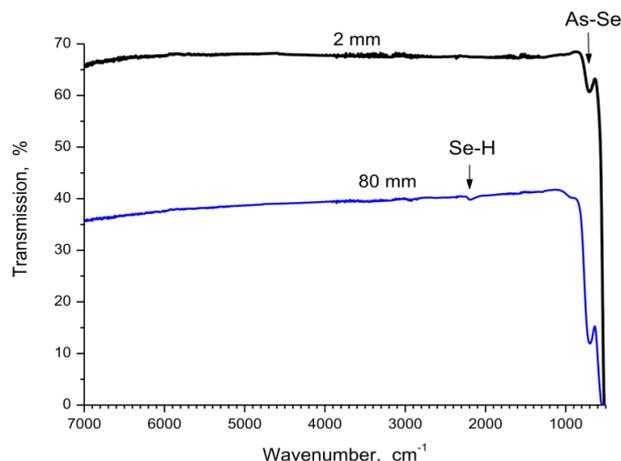

Fig.1. Transmission spectra of $As_{39}Se_{61}$ glass sample with thickness 2 mm and 80 mm.

The estimated content of gas-forming impurities determined from the IR absorption spectra using the known values of extinction coefficients [15, 16] was the following: hydrogen - ≤0.03 ppm(wt) and oxygen - <0.1 ppm(wt), respectively.

Next, three longitudinal holes with a diameter of 2.3 mm around the core with a radius of 260 μm were drilled in the $As_{39}Se_{61}$ glass rod with a diameter of 16 mm and a length of 80 mm. To improve the inner surface of the holes after drilling, of the glass preform was chemically polished in the $H_2SO_4$ + 0,015 mol/l $K_2Cr_2O_7$ solution for 30 min. The cross-section of the rod with 3 holes is shown in Fig. 2 (a).

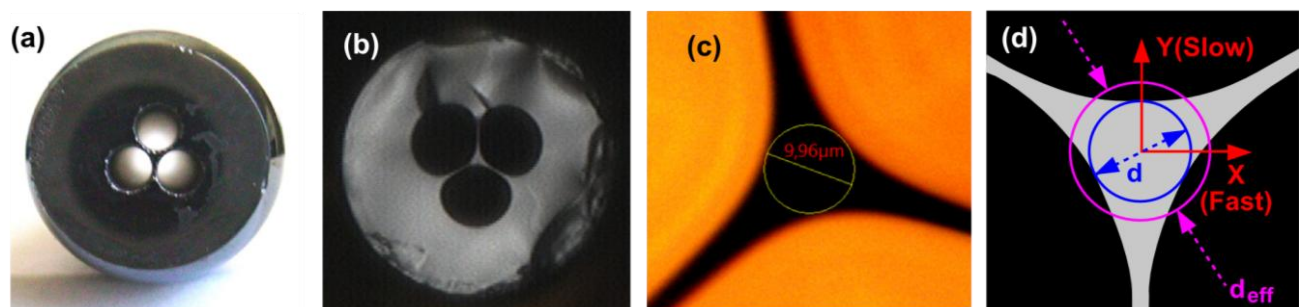

Fig. 2. Cross-section of (a) $As_{39}Se_{61}$ rod with drilled holes, (b, c) SCF with d = 10 μm, (d) modeled SCF for numerical simulations.

Then, the 3-hole preform of high purity $As_{39}Se_{61}$ glass was drawn, using a rod technique, into an SCF with 240 μm outer diameter. The core diameter was 10 μm, the bridge thickness between the holes was 1.3 μm. The SCF cross-section is shown in Fig. 2 (b, c) on different scales. The SCF optical losses of measured using the cut-back technique at a wavelength of 1.06 μm were 5.5±0.5 dB/m.

Further, fiber tapers (bicones) with outer waist diameters of 40-60 μm and core diameters of 1.6-3 μm (and even 0.2-0.6 μm) were stretched from the manufactured pieces of SCF. The waist lengths were about 0.5 cm, the transitional pieces (from a core diameter of 10 μm to a minimal core thickness) were about a few mm. Optical losses of SCF after tapering were not measured due to the too small core diameter.

## *2.2 Theoretical study of SCF parameters*

Before simulation of potential pulse conversion in the prepared SCFs, it was necessary to calculate mode structures, dispersion curves, and other parameters for samples with different core diameters $d$. Hereinafter $d$ is a diameter of the inscribed circle (see Fig. 2 (d)).

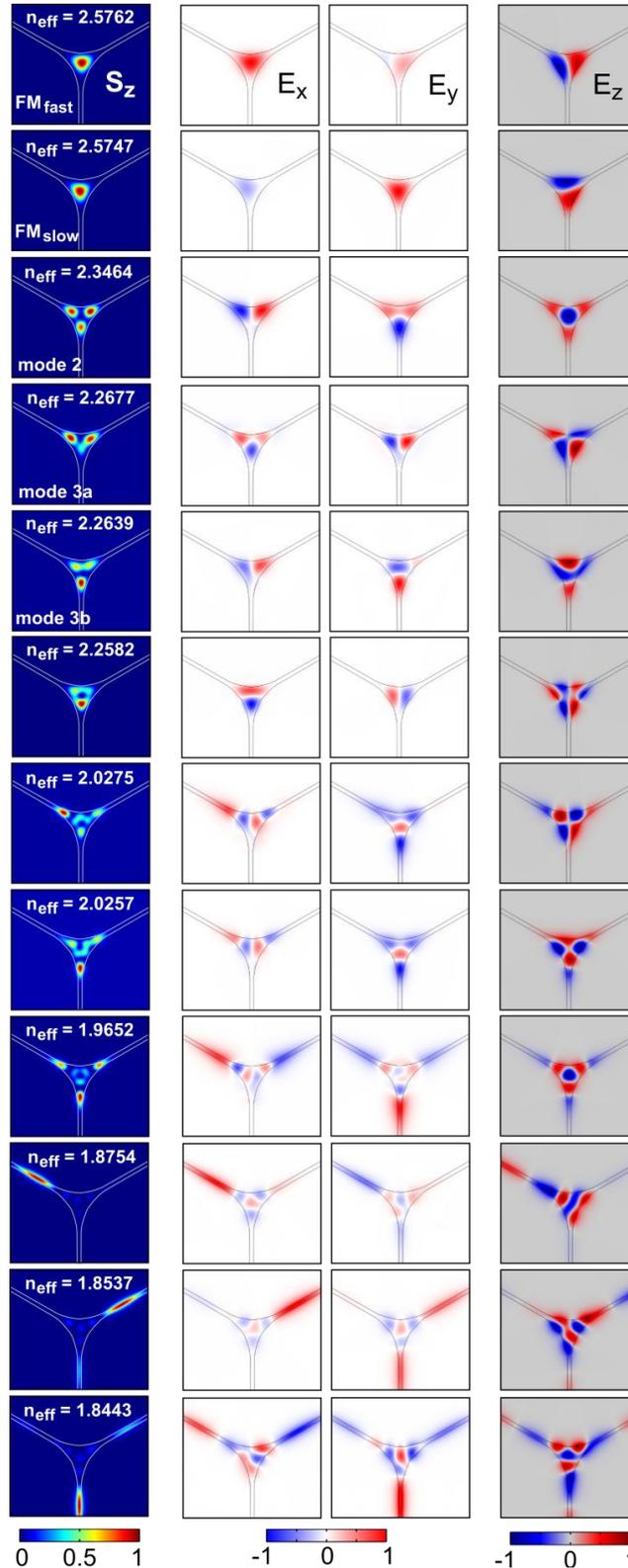

Fig. 3. Distribution of the first 12 modes in SCF with core diameter $d = 1$ μm at a wavelength of 2 μm: longitudinal Poynting vectors (the 1st column), electric field components $E_x$, $E_y$, $E_z$ (the 2nd, 3rd and 4th columns, respectively).

The produced SCFs are multimode even for a very thin core with $d = 1$ μm due to the large index difference between the glass and air. The structure of electric field components ($E_x$, $E_y$, $E_z$) and longitudinal Poynting vector ($S_z$) calculated at the wavelength $\lambda = 2$ μm for several modes sorted in descending order of the effective refractive indexes ($n_{eff}$) are shown in Fig. 3. The calculations were conducted using the finite-element method with the perfectly matched layer boundary conditions. Refractive index measurements of $As_2Se_3$ from Amorphous Materials Inc. [17], which are in a good agreement with the refractive index of $As_{38}Se_{62}$ [18] and, we believe, with our composition $As_{39}Se_{61}$, were taken into account. The fiber cross-section from Fig. 2(d) was used in simulations.

Figure 4 demonstrates the following linear parameters for the first five modes with the largest $n_{eff}$: effective refractive index $n_{eff}$, group delay $\beta_1 = \partial \beta_{eff}/\partial \omega$, where $\beta_{eff}$ is wavenumber and $\omega$ is angular frequency, and dispersion $D = -(2\pi c/\lambda^2)\cdot(\partial^2 \beta_{eff}/\partial \omega^2)$, where $c$ is the speed of light in vacuum.

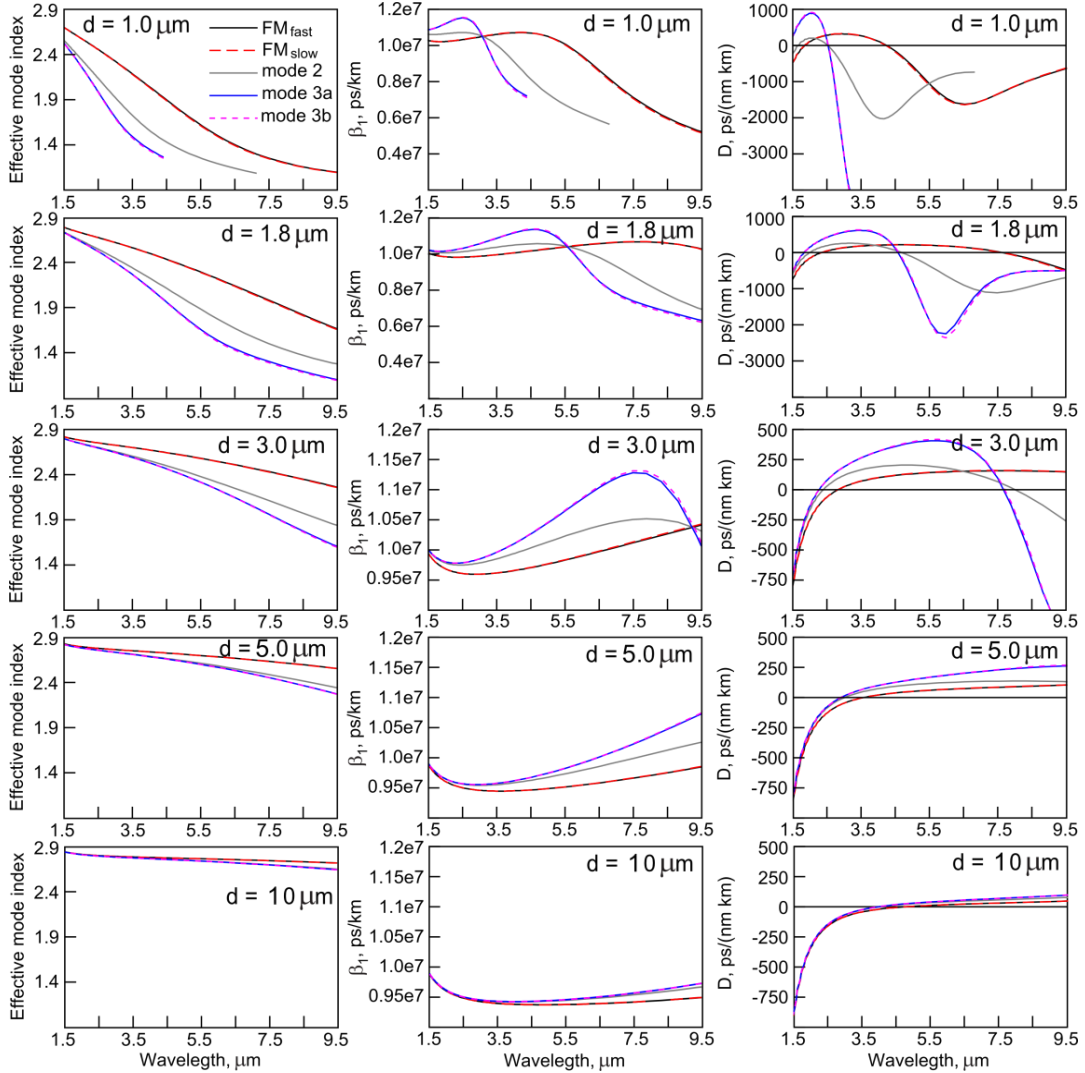

Fig. 4. Calculated linear parameters of the first five modes sorted in descending order of $n_{eff}$ in fibers with different diameters d: effective refractive index (left column); group delay $\beta_1$ (middle column); group-velocity dispersion (right column).

Because of its geometry, the fiber is slightly birefringent, but one can see that the calculated parameters are very similar for the same modes in different polarization (corresponding to modes "$FM_{fast}$", "$FM_{slow}$" and "3a", "3b"). The negligibility of birefringence for 3-holes $As_{38}Se_{62}$ SCF was also noted in [9].

Further, we wonder when one can use the relatively simple approximation of wire waveguides with a round cross-section located in air for dispersion estimation and when this approximation is too rough. The same approach for 6-holes tellurite SCF gave a very good agreement with experimental measurements [19, 20]. So, this simple approximation may be very useful for fiber design. The calculated dispersion parameter D for the FM of the SCF in full-vector model as well as dispersion for the $HE_{11}$ mode of the round wire with the effective diameter $d_{eff}$ ($d_{eff}/d$ is approximately 1.3, see Fig. 2 (d)) using the corresponding characteristic equation [20, 21] are shown in Fig. 5(a) and (b) respectively. Figure 5 (c) demonstrates a comparison between dispersion D-characteristics calculated by two different approaches on magnified scale. One can see that wire approximation gives better results for relatively thick cores when fiber dispersion is closer to the material one. The thinner the core, the shorter wavelengths range of accordance is. However, even for very thin cores, it gives adequate evaluations of the first ZDW. So, this simple estimation may be convenient to design SCF for SC generation, because the ZDW location is of great importance for it [5, 6].

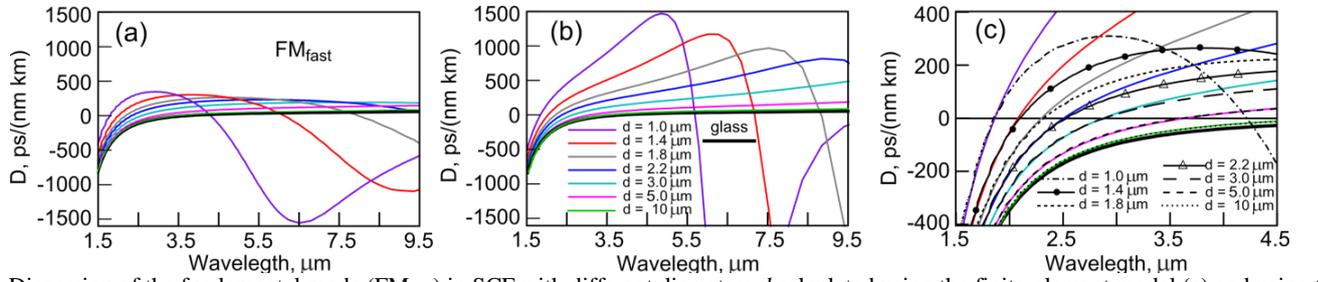
Fig. 5. Dispersion of the fundamental mode (FM$_{fast}$) in SCF with different diameters $d$ calculated using the finite-element model (a) and using the characteristic equation for HE$_{11}$ mode of the round wire (b). Dispersion calculated by two mentioned above approaches on magnified scale (c).

The next step was to calculate effective FM area $A_{eff}(\omega)$ and energy fraction $g(\omega)$ in the glass-core (not in the air holes) for various diameters (Fig.6 (a) and (b)). These parameters have impact on the nonlinearity coefficient. The smaller $A_{eff}$ and the higher $g(\omega)$, the higher nonlinearity is.

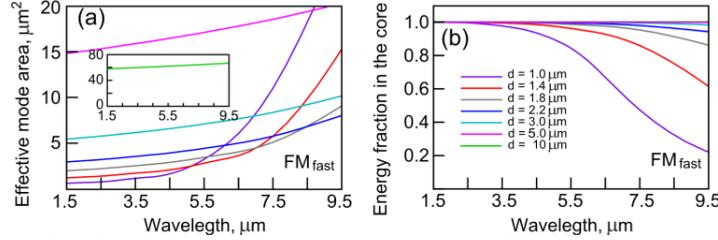
Fig. 6. (a) The effective field area $A_{eff}$ of the fundamental mode (FM$_{fast}$) in SCF with different diameters $d$. The inset shows the area of FM$_{fast}$ for $d = 10$ μm. (b) The ratio of the energy propagating in the core to the total energy $g$ of FM$_{fast}$.

## 2.3 Numerical simulation of nonlinear pulse dynamics in SCF

To simulate nonlinear dynamics of ultrashort pulses propagating in the fundamental mode in a tapered SCF, we use the generalized nonlinear Schrödinger equation with the calculated dispersion taken into account (shown in Fig. 5(a)), the third-order instantaneous Kerr and retarded Raman nonlinearities, linear loss $\alpha(\omega)$, as well as frequency dependence of effective mode field area (shown in Fig. 6), and with the addition of two-photon absorption function $\alpha_{TPA}(z, \omega)$ as in the paper [22]. For a frequently assumed mid-IR pump, two-photon absorption can be neglected, but in our case of near-IR pumps, this process impacts noticeably on nonlinear dynamics. The propagation equation for a signal envelope launched at central frequency $\omega_0$ was the following [23,24]:

$$\frac{\partial \bar{A}'(z,\omega)}{\partial z} = \exp\left[-\hat{L}(z,\omega)z\right] \cdot \left[i\bar{\gamma}(z,\omega) - \alpha_{TPA}(z,\omega)\right] \cdot \hat{F}\left\{A(z,t)\int_{-\infty}^{\infty} R(T')\left|A(z,t-T')\right|^2 dT'\right\}$$

where $z$ is the distance of propagation, $t$ is the time in the retarded system, F and $F^{-1}$ are the operators of Fourier and inverse Fourier transforms,

$$A(z,t) = F^{-1}\left\{\frac{\bar{A}(z,\omega)}{A_{eff}^{1/4}(z,\omega)}\right\},$$

$$\bar{A}'(z,\omega) = \bar{A}(z,\omega)\exp\left[-\hat{L}(z,\omega)z\right]$$

and the linear operator contains fiber dispersion and linear loss:

$$\hat{L}(z,\omega) = i\left[\beta(z,\omega) - \beta(z,\omega_0) - (\omega-\omega_0)\cdot\beta_1(z,\omega_0)\right] - \alpha(\omega)/2.$$

The frequency dependent nonlinear coefficient is defined as

$$\bar{\gamma} = \frac{\omega n_2 n_0 g(\omega)}{c n_{eff}(z,\omega) A_{eff}^{1/4}(z,\omega)},$$

where $n_2 = 2.2 \cdot 10^{-17}$ m$^2$/W is the nonlinear refractive index, and $n_0$ is the linear refractive index at the wavelength where $n_2$ is determined.

The retarded Raman response function is

$$R(t) = (1-f_R)\delta(t) + f_R h_R,$$

$$h_R(t) = f_a \tau_1 (\tau_1^{-2} + \tau_2^{-2})\exp(-\tau/\tau_2)\sin(\tau/\tau_1) + f_b \tau_3(\tau_3^{-2} + \tau_4^{-2})\exp(-\tau/\tau_4)\sin(\tau/\tau_3),$$

$f_R = 0.1$, $f_a = 0.7$, $f_b = 0.3$, $\tau_1 = 23$ fs, $\tau_2 = 230$ fs, $\tau_3 = 20.5$ fs, and $\tau_4 = 260$ fs [24].

Figure 7(a) contains the modeled profile of a core diameter $d$ vs $z$ and Fig. 7(b) shows the corresponding ZDWs dependence on $z$. Note, there are two ZDWz for thin cores (ZDW$_1$ and ZDW$_2$) with anomalous dispersion between them. For wavelengths shorter than ZDW$_1$ and longer than ZDW$_2$, the dispersion is normal. We considered ultrashort pump pulses at 1.57 and 2 μm, which can be easily produced with Er: fiber and Tm: fiber standard techniques [25-27]. The duration of transform-limited hyperbolic secant pump pulses was 150 fs (full width at half maximum). The numerically simulated spectral evolution of 50-pJ pulses calculated by the split-step (Fourier) method using fast Fourier transform, is presented in Fig. 7(c) and (d) respectively for pump wavelengths of 1.57 and 2 µm. Spectra at the output of a tapered SCF for different pump pulse energies are shown in Fig. 8 for a pump wavelength of 1.57 µm and in Fig. 9 for a pump wavelength of 2 µm.

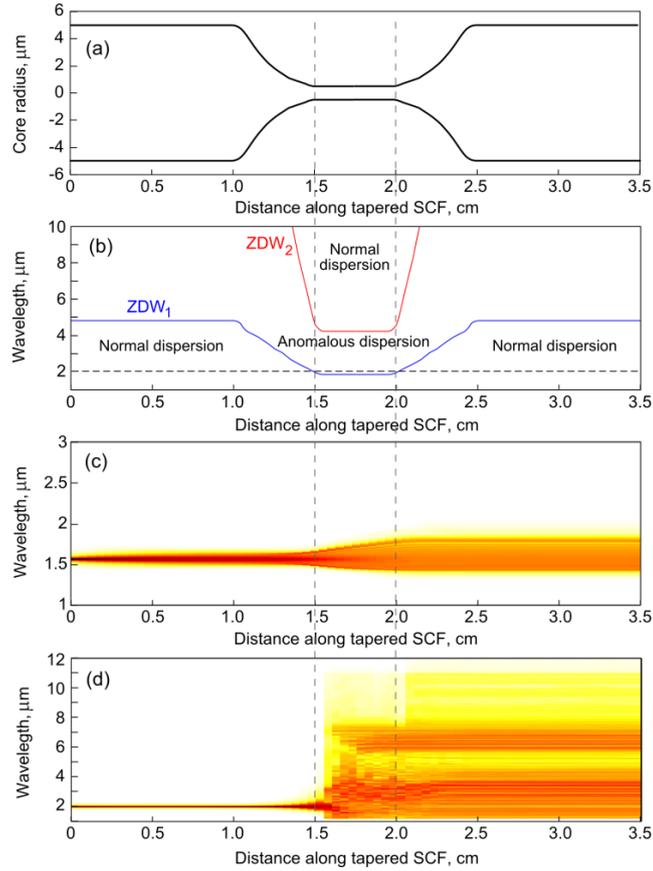

Fig. 7. Modeled profile of core diameter $d$ vs $z$ (a), corresponding $ZDW_1$ (blue) and $ZDW_2$ (red) dependences on $z$ (b), numerically simulated spectral evolution of 50-pJ pulses at 1.57 (c) and 2 µm (d).

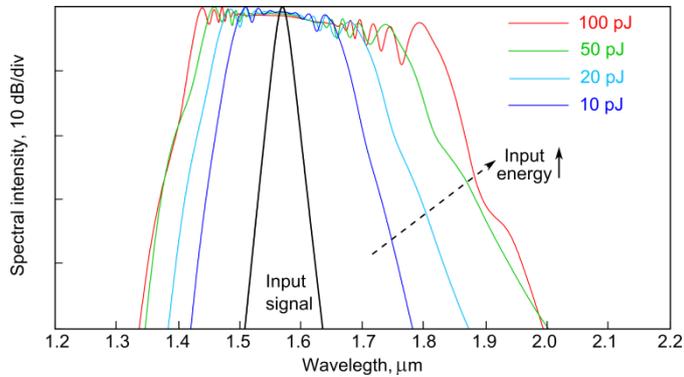

Fig. 8. Simulated spectra at the output of As-Se tapered SCF with profile shown in Fig. 7(a) pumped at 1.57 µm with different pulse energies.

### *2.4 Experimental study of spectral broadening in SCF pumped at 1.57 µm*

The experiments were carried out with a pump source at 1.57 µm, previously implemented for SC generation in germanate fibers [28]. Here, for a study of the nonlinear spectral broadening in the produced As-Se tapered SCF, we also used a femtosecond Er-doped fiber laser system consisting of a master oscillator with a pulse repetition rate of about 50 MHz, a Faraday isolator, an amplifier and one more Faraday isolator to prevent gain decreasing due to back reflections from the As-Se fiber cleave. The diode-pumped at 975 nm master oscillator was passively mode-locked using a nonlinear rotation of the polarization ellipse of a femtosecond pulse due to the optical Kerr effect. The pulses from the output of the Er: fiber laser system were launched by using a focusing lens into the As-Se tapered SCF with the input core diameter of 10 µm (for the initial segment of about 1 cm) and the waist core diameter of about 1.5 µm. The coupled efficiency was estimated to be about 10%. The scheme of the experimental setup is presented in Fig. 10 (a). The spectra of the input signal and the converted pulses at the tapered SCF are shown in Fig. 10 (b). The spectral resolution was 3±0.1 nm. The error is determined by the accuracy of the positioning of the gratings. We smoothly enlarged the power and found that the higher the power, the broader the output spectrum is. However, the spectral width was limited by SCF damage: a bridge was burned at a power exceeding 50 mW. The cross-sections of the damaged and intact SCFs are shown in Fig 10 (c, d).

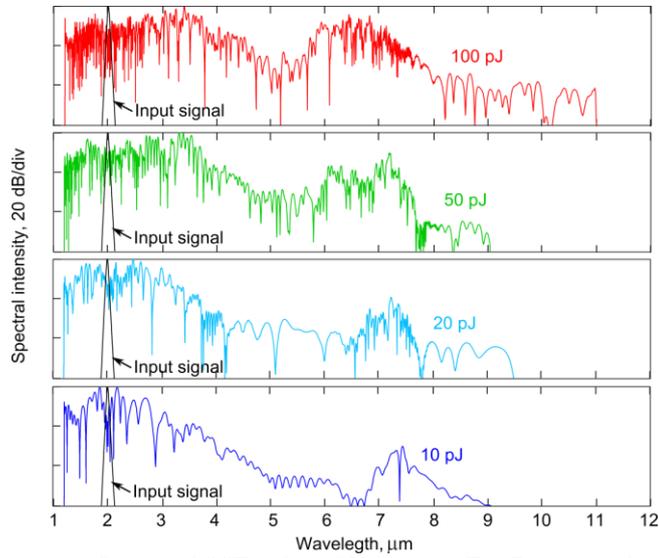

Fig. 9. Simulated spectra at the output of As-Se tapered SCF with profile shown in Fig. 7(a) pumped at 2 μm with different pulse energies.

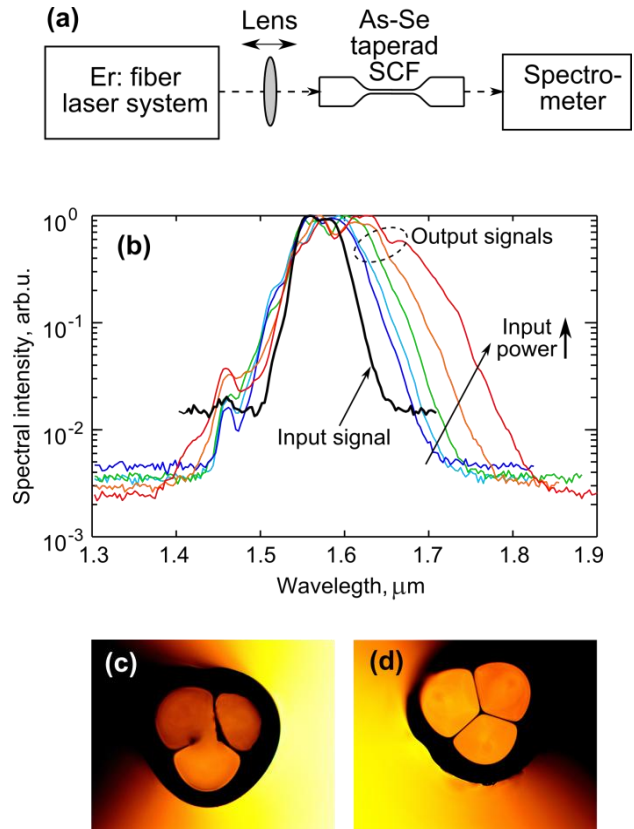

Fig. 10. (a) Experimental setup. (b) Experimental spectra at the output of As-Se tapered SCF with input core diameter of 10 μm and waist core diameter of 2μm for different (unmeasured exactly) input powers. Cross-section of damaged (c) and initial SCF (d).

## 3 Discussion

We designed and manufactured a tapered SCF of high-purity $As_{39}Se_{61}$ glass for wavelength conversion of ultrashort signals from standard *near-IR* fiber laser sources to the mid-IR. We presented experimental and theoretical study of produced samples parameters: transparencies, fiber geometrical structures, mode fields, effective refractive indices, group delays between signals in different modes, dispersion for several lowest modes, effective mode area of the fundamental mode, and the ratio of the energy propagating in the core to the total energy.

It was shown that the thinner the fiber core, the shorter the ZDW is. For a core diameter $d$ of 1 μm, the first ZDW is 1.85 μm but for $d = 1.4$ μm, ZDW is 2.1 μm. So, for operating in the anomalous dispersion regime, the core diameter of the fiber waist should be about 1 μm with a pump at 2 μm, but this regime is impossible with a pump at 1.57 μm for the considered parameters (see Fig. 5). So, it seems obvious that for mid-IR SC generation, pump at 2 μm and a core diameter of a fiber waist of about 1 μm are desirable. To verify this statement, numerical simulation of nonlinear pulse dynamics based on generalized nonlinear Schrödinger equation was carried out taking into account the actual dispersion profile, third-order instantaneous Kerr and retarded Raman nonlinearities, linear loss, as well as the frequency dependence of effective mode field area with the addition of two-photon absorption. Although, as one can see in Fig. 3, the fibers are multimode at the considered pump wavelengths of 1.57 and 2 μm, we modeled pulse propagation only for the fundamental mode in accordance with the work [29] demonstrating that, if the pump pulse is short enough ($\leq$ 10ps), higher order modes are not important for SC bandwidth because of temporal walk-off; no light is transferred between the modes [29]. However, the data presented in Figs. 3, 4 may be useful for multimode simulation of SC generation.

As is seen in Fig. 7, the main spectral broadening occurs in a fiber waist with favorable dispersion and nonlinearity. Note that using a non-tapered thin SCF is challenging due to a very low pump coupling efficiency but the tapered profile provides a reasonable value.

The broadband conversion of pump wavelength of 1.57 μm was studied in the experiment. The experimental spectra in Fig. 10 (b) are consistent with the ones simulated numerically and presented in Fig. 8. It was found theoretically that the spectral width is limited by all-normal dispersion, but the experimental limit is related to sample damage. The low experimental damage threshold at 1.57 μm may be associated with a high impact of two-photon absorption. SCFs were multimode and a burned bridge may indicate, as we believe, that the fiber mode in Fig. 3 corresponding to $n_{eff}$ of 1.8754, 1.8537 or 1.8443 could be excited. It could lead to a tightly focused pump in a very small area and a pump intensity exceeding the damage threshold.

Using a pump source at 2 μm significantly reduces two-photon absorption [22] and allows operating in anomalous dispersion regime. One can see in Fig. 7 and 8 that the wavelength of the SC red boundary is shorter than 2 μm for the pump at 1.57 μm. But with a pump at 2 μm for the same energies, the SC red boundary is located in the mid-IR and even beyond 10 μm for the coupled energy of 100 pJ (see Fig. 7, 9). In a fiber waist, SC generation is dominated by soliton dynamics (with a pump at 2 μm) [5, 30]. It was revealed theoretically that the signals generated in the fiber waist with maximum spectral intensity at the wavelengths near 7 μm (see Fig. 7(d), 9) are red-shifted dispersive waves beyond the second ZDW ($ZDW_2$ in Fig. 7(b)) [20, 31]. A transitional fiber piece after the waist also contributes to generating long-wavelengths component because its dispersion becomes anomalous (see Fig. 5(a) and 7 (b)) and nonlinearity increases (see Fig. 6) for these dispersive waves near 7 μm (produced in the fiber waist).

## 4 Conclusion

We developed a tapered SCF of high-purity $As_{39}Se_{61}$ glass for SC generation in the mid-IR pumped by a standard *near-IR* fiber laser source at 2 μm. The samples parameters were studied experimentally and theoretically. Microstructuring allowed shifting the ZDW to the range shorter than 2 μm in the fiber waist for operating in the anomalous dispersion regime. SC generation with red boundary beyond 10 μm was obtained numerically with ultrashort pump pulses. However, we showed that a pump at 1.57 μm is not suitable for the mid-IR SC generation due to all-normal dispersion and experimentally observed low damage threshold. The proposed design of a tapered SCF using an ultrashort pump at 2 μm may be very promising for SC source development. This simple concept may rival with schemes using ChG fibers pumped by complicated mid-IR optical parametric systems.

**Acknowledgments**
We thank Dr. Arkady Kim for comprehensive support and numerous fruitful discussions, Prof. Gennady Snopatin for fiber drawing, and Dr. Alexey Andrianov for helpful experimental advice.
The experimental work is supported by the Russian Science Foundation (Grant No. 16-13-10251). The numerical simulation provided by E.A. Anashkina is supported by the Russian Foundation for Basic Research (Grant No 16-32-60053).

**References**
[1]	G. Tao, H. Ebendorff-Heidepriem, A.M. Stolyarov, S. Danto, J.V. Badding, Y. Fink, J. Ballato, A.F. Abouraddy, Infrared fibers, Adv. Opt. Photon. 7 (2015) 379-458. https://doi.org/10.1364/AOP.7.000379.
[2]	K. Yin, B. Zhang, L. Yang, and J. Hou, 15.2 W spectrally flat all-fiber supercontinuum laser source with >1 W power beyond 3.8 μm, Opt. Lett. 42 (2017) 2334-2337. https://doi.org/10.1364/OL.42.002334.
[3]	S. Kedenburg, C. Strutynski, B. Kibler, P. Froidevaux, F. Désévédavy, G. Gadret, J.-C. Jules, T. Steinle, F. Mörz, A. Steinmann, H. Giessen, and F. Smektala, High repetition rate mid-infrared supercontinuum generation from 1.3 to 5.3 μm in robust step-index tellurite fibers, J. Opt. Soc. Am. B 34 (2017), 601-607. https://doi.org/10.1364/JOSAB.34.000601.
[4]	Z. Zhao, B. Wu, X. Wang, Z. Pan, Z. Liu, P. Zhang, X. Shen, Q. Nie, S. Dai, R. Wang, Mid-infrared supercontinuum covering 2.0–16 μm in a low-loss telluride single-mode fiber, Laser & Photonics Reviews 11 (2017) 1700005. doi: 10.1002/lpor.201700005.
[5]	J. M. Dudley, G. Genty, and S. Coen, "Supercontinuum generation in photonic crystal fiber," Rev. Mod. Phys. 78, 1135–1184 (2006). https://doi.org/10.1103/RevModPhys.78.1135.
[6]	M. Frosz, P. Falk, and O. Bang, "The role of the second zero-dispersion wavelength in generation of supercontinua and bright-bright soliton-pairs across the zero-dispersion wavelength," Opt. Express 13 (2005), 6181–6192. https://doi.org/10.1364/OPEX.13.006181.
[7]	C.R. Petersen, U. Møller, I. Kubat, B. Zhou, S. Dupont, J. Ramsay, T. Benson, S. Sujecki, N. Abdel-Moneim, Z. Tang, D. Furniss, A. Seddon, O. Bang, Mid-infrared supercontinuum covering the 1.4-13.3 μm molecular fingerprint region using ultra-high NA chalcogenide step-index fibre, Nat. Photonics 8 (2014), 830-834. doi:10.1038/nphoton.2014.213.
[8]	T. Cheng, K. Nagasaka, T.H. Tuan, X. Xue, M. Matsumoto, H. Tezuka, T. Suzuki, Y. Ohishi, Mid-infrared supercontinuum generation spanning 2.0 to 15.1 μm in a chalcogenide step-index fiber, Opt. Lett. 41 (2016) 2117-2120. https://doi.org/10.1364/OL.41.002117.
[9]	U. Møller, Y. Yu, I. Kubat, C.R. Petersen, X. Gai, L. Brilland, D. Méchin, C. Caillaud, J. Troles, B.Luther-Davies, O. Bang, Multi-milliwatt mid-infrared supercontinuum generation in a suspended core chalcogenide fiber, Opt. Express 23 (2015) 3282-3291. https://doi.org/10.1364/OE.23.003282.
[10]	M.F. Churbanov, V.S. Shiryaev, Crystallization of chalcogenide glasses, Vysokochistye Veschestva (High-Purity Substances) 4 (1994) 21-33 (in Russian)
[11]	V.S. Shiryaev, M.F. Churbanov, G.E. Snopatin, F. Chenard, Preparation of low-loss core–clad As–Se glass fibers, Opt. Mater., 48 (2015) 222–225. http://dx.doi.org/10.1016/j.optmat.2015.08.004.
[12]	V.S. Shiryaev, M.F. Churbanov, E.M. Dianov, V.G. Plotnichenko, J.-L. Adam, J. Lucas, Recent progress in preparation of chalcogenide As-Se-Te glasses with low impurity content, J. Optoelectron. Adv. Mater. 7 (2005) 1773-1780.

[13]   V.S. Shiryaev, J.-L. Adam, X.H. Zhang and M.F. Churbanov, Study of characteristic temperatures and nonisothermal crystallization kinetics in As-Se-Te glass system, Solid State Sci. 7 (2005) 209-215. http://dx.doi.org/10.1016/j.solidstatesciences.2004.10.027
[14]   M.F. Churbanov, V.S. Shiryaev, S.V. Smetanin, E.M. Dianov, V.G. Plotnichenko, Q. Hua, G. Li, and H. Shao, Effect of sulfur on the optical transmission of $As_2Se_3$ and $As_2Se_{1.5}Te_{1.5}$ glasses in the range 500-1100 $cm^{-1}$. Inorg. Mater. 35 (1999) 1229-1234.
[15]   G.E. Snopatin, V.S. Shiryaev, G.E. Plotnichenko, E.M. Dianov, M.F. Churbanov, High-purity chalcogenide glasses for fiber optics, Inorg. Mater. 45(13) (2009) 1439-1460.
[16]   V.S. Shiryaev, S.V. Smetanin, D.K. Ovchinnikov, M.F. Churbanov, E.B. Kryukova, and V.G. Plotnichenko, Inorg. Mater. 41 (2005) 308–314. doi:10.1134/S0020168509130019.
[17]   Amorphous Materials Inc., "AMTIR-2," (2014). Online at http://www.amorphousmaterials.com/.
[18]   C. Caillaud, G. Renversez, L. Brilland, D. Mechin, L. Calvez, J.-L. Adam, and J. Troles, Photonic Bandgap Propagation in All-Solid Chalcogenide Microstructured Optical Fibers, Materials 7, 6120–6129 (2014). doi:10.3390/ma7096120.
[19]   M.Yu. Koptev, E.A. Anashkina, A.V. Andrianov, V.V. Dorofeev, A.F. Kosolapov, S.V. Muravyev, A.V. Kim, Widely tunable mid-infrared fiber laser source based on soliton self-frequency shift in microstructured tellurite fiber, Opt. Lett. 40 (2015) 4094-4097. https://doi.org/10.1364/OL.40.004094.
[20]   E.A. Anashkina, A.V. Andrianov, V.V. Dorofeev, A.V. Kim, Toward a mid-infrared femtosecond laser system with suspended-core tungstate-tellurite glass fibers, Appl. Optics 55 (2016) 4522-4530. https://doi.org/10.1364/AO.55.004522.
[21]   C. Chaudhari, T. Suzuki, Y. Ohishi, Design of Zero Chromatic Dispersion Chalcogenide $As_2S_3$ Glass Nanofibers, J. Lightwave Technol. 27 (2009), 2095-2099. doi: 10.1109/JLT.2008.2007223
[22]   A. Al-kadry, C. Baker, M. El Amraoui, Y. Messaddeq, M. Rochette, Broadband supercontinuum generation in $As_2Se_3$ chalcogenide wires by avoiding the two-photon absorption effects, Opt. Lett. 38 (2013) 1185-1187. https://doi.org/10.1364/OL.38.001185
[23]   J. Laegsgaard, Mode profile dispersion in the generalized nonlinear Schrödinger equation, Opt. Express 15 (2007), 16110-16123. https://doi.org/10.1364/OE.15.016110.
[24]   W. Yuan, 2–10 μm mid-infrared supercontinuum generation in $As_2Se_3$ photonic crystal fiber, Laser Phys. Lett. 10 (2013) 095107. DOI: 10.1088/1612-2011/10/9/095107.
[25]   A.V. Andrianov, E.A. Anashkina, S.V. Muravyev, A.V. Kim, Hybrid Er/Yb fibre laser system for generating few-cycle 1.6 to 2.0 μm pulses optically synchronised with high-power pulses near 1 μm", Quantum Electronics, 43(2013), 256–262. https://doi.org/10.1070/QE2013v043n03ABEH015132.
[26]   E.A. Anashkina, A.V. Andrianov, M.Yu. Koptev, S.V. Muravyev, A.V. Kim, Towards Mid-Infrared Supercontinuum Generation with Germano-Silicate Fibers, IEEE J. Sel. Top. Quantum Electron. 20 (2014) 643-650. http://dx.doi.org/10.1109/JSTQE.2014.2321286.
[27]   M.Yu. Koptev, E.A. Anashkina, A.V. Andrianov, S.V. Muravyev, A.V. Kim, Two-color optically synchronized ultrashort pulses in Tm/Yb co-doped fiber amplifier, Opt. Lett. 39 (2014) 2008—2011. https://doi.org/10.1364/OL.39.002008.
[28]   E.A. Anashkina, A.V. Andrianov, M.Yu. Koptev, V.M. Mashinsky, S.V. Muravyev, A.V. Kim, Generating tunable optical pulses over the ultrabroad range of 1.6-2.5 μm in GeO2-doped silica fibers with an Er:fiber laser source, Opt. Express 20 (2012), 27102-27107. https://doi.org/10.1364/OE.20.027102
[29]   I. Kubat, O. Bang, Multimode supercontinuum generation in chalcogenide glass fibres, Opt. Express 24 (2016) 2513-2526. https://doi.org/10.1364/OE.24.002513.
[30]   E.A. Anashkina, A.V. Andrianov, S.V. Muravyev, and A.V. Kim, All-fiber design of Erbium-doped laser system for tunable two-cycle pulse generation, Opt. Express 19 (2011), 20141-20150. https://doi.org/10.1364/OE.19.020141.
[31]   T. Cheng, T.H. Tuan, L. Liu, X. Xue, M. Matsumoto, H. Tezuka, T. Suzuki, Y. Ohishi, Fabrication of all-solid $AsSe_2$-$As_2S_5$ microstructured optical fiber with two zero-dispersion wavelengths for generation of mid-infrared dispersive waves, Applied Physics Express 9 (2016) 022502. https://doi.org/10.7567/APEX.9.022502.